# Next Generation Quantum Dots Based Multijunction Photovoltaics

[1*]Ankul Prajapati, [2]Bade M. H.

[1,2] Department of Mechanical Engineering, Sardar Vallabhbhai National Institute of Technology, Surat, Gujarat, 395007, India

*ankulprajapati7@gmail.com

**Abstract**

Photovoltaic cells (PVc), as an energy provider to the next generation and the biggest source of renewable energy. Since the last decade improving efficiency and reducing the cost of PVc has been a subject of active research among scientists. Promising progress in the field of material science and manufacturing process at Nano-level played a big role. Still, at present there are many challenges before photovoltaics for efficient and economic energy. However, Photovoltaics cell based on p-n type homojunction semiconductors with different organic and inorganic materials reported thus for generally suffer from poor performance. According to the available literature, colloidal quantum dots having immense properties like a wide range of light absorption, easily charge separation and transport. To utilize the maximum part of the spectrum of solar energy reaching to the earth and making effective energy production, here we introduce the complete cell architecture and numerical investigation on quantum dot based solar cells (QDSCs) with a heterostructure multijunction approach. Successive ionic layer adsorption at different heterogeneous interfaces were analyzed. We majorly focused on improving the electrical and optical properties of the QDSCs achieved by different materials and structural approaches. Here, we report a heterostructure II-Type of band alignment engineering strategy for QDSCs interfaces that significantly enhances the efficiency descriptors. In the context of intermediate band solar cell (IBSC), we investigated optical properties of QDs and strain effects on multilayer PVc and we summarize the strain effect in QDs growth and local energy band bending of conduction band (CB) and valence band (VB).

## I. Introduction

Quantum material always has been actively involved in the advancement of scientific instrument and engineering components [1-4]. Exploring the power of the single atom has been the topic of an ever-increasing interest among researchers. Promising progress in quantum dots and making the use of extraordinary properties of quantum dots (QDs) [5-9] giving the next generation efficient and compact semiconductor devices [10-11]. Among the energy sector, extensively using the quantum dots for making third generation thin film highly efficient solar cell. As we know that solar cell is playing a big role as a green and renewable energy source [12]. At the same time, there are major troubles like highly expensive and large area requirement for a small amount of energy production. In present, almost the market is dominated by second-generation solar cell based on homo-junction p-n type inorganic materials such as Si and Ge [13,14]. Recently some PVc has been reported based on organic and perovskite materials but less stable. Second generation solar cell is cheaper than first generation single junction crystalline solar cell, but not efficient more than 16% [15]. So, what are the requirements for making photovoltaics cell as an economic, efficient and reliable source of energy? There are a number of papers addressing this problem and proposed different ways of making more efficient solar cells [16-21]. Since last five years major studies are focused on colloidal quantum dots-based PVc. This is only because of their engrossing properties and functionalities, many of which depend in a spectacular way on their Nano size [22]. Here we could analyze the size and shape factor of QDs and discuss effects of induced strain in the potential energy of conduction band (CB) and valence band (VB) of the InAs quantum dot. Effective masses of electrons and holes of materials used in PVc have been considered for such electronic parameter changes. After the first successful synthesis of colloidal CdS/Se/Te QDs by Murray et al [23] it is the most studied QDs due to their exceptional optical and electrical properties ever seen in semiconductors. By the time other alloy semiconductors came in the picture even with more powerful features. Semiconducting colloidal quantum dots (CQDs) such as InAsN, InPN, InAsSb, PbS, GaAs and InAs captured the field of optoelectronics research and quantum technology. Particularly photonics devices such as photovoltaics cell, photo sensors, single photon detector, and quantum circuits. We have been analyzed heterostructure interface between the successive layer of QDs based multijunction solar cell. Though heterojunction between different semiconductors has been extensively studied in the literature [24-28]. And also, there is enough number of experimental demonstrations of the nobility of heterogeneous technology. By using the same methodology scientists already have been developed more compact transistors and integrated circuits [29,30]. Still, in photovoltaics application there is no such commercial and stable devices. There is to some extent but limited to space applications due to their high cost. However, the efficient exploitation of the colloidal quantum dots towards high-performance PV cells, it is important to understand intermediate band of multijunction cell and wetting layer matrix composition on the evolution of QDs shape, size, electronic structure, and optical orientation. In this context, we report on the numerical investigation on electron confinement inside a single quantum dot, electron-hole pair generation, multiple intermediate energy band gap and absorption of the maximum solar spectrum to harness solar photon energy in energy conversion. We have analyzed falling solar ray's angle with respect to vertical & horizontal plane and percentage of absorption, results were analyzed and plots for the same study have been presented. Recently encountered problems on the material selection for QDSC are also presented and along with proposed solutions. Ultimately this paper aims to actively encourage further investigation and development in QDs based hetero-structured multi-junction photovoltaic cells.

## II. Basic principles of Multilayered QDs Based PVc

Colloidal quantum dots are a semiconductor nanocrystal with unique properties confined in three-dimensional space. QDs are also having two dimensional and one-dimensional confinement. Two factors were majorly involved in describing the quantum confinement of QDs in bulk Nanocrystals. First, de Broglie wavelength and second, the Bohr radius. According to de Broglie wavelength of electrons and holes in the bulk semiconductor, $\lambda_e/\lambda_h$ is given as follows,

$$\lambda_e/\lambda_h = \frac{h}{m_{eff}kt} \quad (1)$$

Where $m_{eff}$ is the effective mass of the electron and hole respectively. When one dimension of the Nanocrystal approach to the $\lambda_e/\lambda_h$, quantum confinement effects begin to be important [30]. For most of the semiconductors, it ranges from 5 to 100 nm. The Bohr radius ($R_b$), which explain the characteristic of a photogenerated electron-hole pairs in Nano-crystal, is evaluated as,

$$R_b = \frac{\varepsilon h^2}{\mu e^2} \quad (2)$$

where $\varepsilon$ is the dielectric constant of the semiconductor and μ is the reduced effective mass of the electron and hole. Generally, it ranges from 2 to 100 nm [31]. Since 2002, the first deployment of QDs in solar cell by Nozik, semiconducting materials advancement goes on. The first time he introduced mainly three general strategies to integrate the QDs into PVc application, first use the QDs to sensitize wide bandgap semiconductors, second, place the QDs array in intimate contact with charge conducting polymers and third, allow the efficient electron/hole conductivity from QDs arrays, where the QDs are electronically coupled [32]. In earlier QDs PVc there were some problems, such as low carrier mobility in QDs film. To eliminate this from PVc intimate connection with a charge-separating interface was required. By the time concepts improved and quantum dots based solar cell got some appreciable efficiency. Along with easy separation and transportation of electron and hole pairs from bulk Nanocrystals, other optoelectronic properties of QDs such as interparticle interaction, thin film morphology, and Nanocrystal lattice chemistry work as efficiency descriptor of QDs based PVc [33]. We have analyzed the quantum confinement and plotted the eigenstates at a different energy level.

When we talk about the quantum mechanical description of charge carriers in multilayered QDs solar cell, detailed explanation of their wave function and dynamic growth become more important. Wave functions of charge carriers, $\Psi_{n,k}(x)$, are found by solving the Schrödinger equations with single electron approximation,

$$H_0\psi_{n,k}(x) = E_{n,k}\psi_{n,k}(x) \quad (3)$$

The Hamiltonian $H_0$ in Eq-3 is given as follows,

$$H_0 = \frac{p^2}{2m_0} + V(x) \quad (4)$$

Where p is quantum mechanical momentum operator, defined as, $p = -i\hbar\nabla$ and $V(x)$ is the crystal potential experienced by charge carriers, which is a periodic function, with the period of X of Nanocrystal lattice, $V(x) = V(x + X)$. Here we have adopted $k.p$ perturbation theory method, which is an approximated semi-empirical approach for evaluating the band structure and optical properties of QDs in bulk Nanocrystalline solids [34-37]. By Bloch's theorem, the solutions of Schrödinger equation (3), can be written as follows,

$$\psi_{n,k}(x) = e^{ikx}u_{n,k}(x) \quad (5)$$

where k is wave vector of charge carriers, n is the discrete index also known as band index, and $u_{n,k}$ is the periodic function with the same periodicity X as the Nanocrystal lattice. The periodic function, $u_{n,k}$ satisfies Eq-3, which can be written as follows,

$$H_k u_{n,k}(x) = E_{n,k}u_{n,k}(x) \quad (6)$$

Where the Hamiltonian ($H_k$) is given as follows,

$$H_k = \frac{p^2}{2m_0} + V(x) + \frac{\hbar^2 k^2}{2m_0} + \frac{\hbar k.p}{m_0} \quad (7)$$

Eq. (7) can be presented as the sum of two terms, $H_0$ and $H'_k$, $H_k = H_0 + H'_k$. Where $H_0$ is the unperturbed Hamiltonian, which in fact equals the exact Hamiltonian at $k=0$ and $H'_k$ is the perturbation Hamiltonian is also called as $k.p$ Hamiltonian. As we can observe that, the perturbation term $H'_k$ gets progressively smaller as value of $k$ tends to zero. It clears that, the $k.p$ perturbation theory gives the most accurate results for small value of $k$ [38,39].

## III. PVc Architecture and Material Selection

QDs heterostructure has been suggested as the best possible materials for highly efficient multilayer-structured PVc. Production of pairs of photogenerated charge carriers (electron-hole) and their diffusive transport via efficient separation at the type-II heterointerface between QDs and wetting layer is sufficient to harness the solar energy. In the proposed complete architecture of PVc (as shown in fig (1)), QDs is made of InAs while substrate, wetting and capping layer is made of AlAs, InAs and GaAs respectively. All three are of zinc blende structure and considered as a single value of crystal lattice constant equal to 0.5nm. All dimensions of proposed PVc architecture is summarized in tab.1. Based on the previously reported experimental data regarding QDs geometry, we proposed a pyramid geometry with a square base for this analysis.

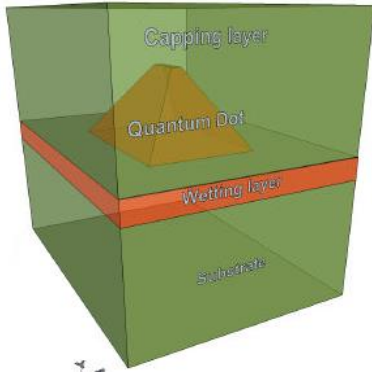

Fig. 1: Architecture of QDs based multi-layered PVc

Tab. 1: Dimensions of different layers

| Dir. | Sub. L | WL | QD | Cap. L |
|---|---|---|---|---|
| x | 10nm | 10nm | 5nm | 10nm |
| y | 10nm | 10nm | 5nm | 10nm |
| z | 5nm | 1nm | 3nm | 6nm |

We investigated different constraint for material selection of QDs PVc, such as optimum band gap and negligible valence band offset. Both of these factors are highly dependent on photon-material interactions and lattice mismatch between the substrate layer and QDs. Which introduces elastic strain in the structure during epitaxial growth of QDs. And causes a significant change in band alignment and eigenstate of QDs materials. We have used lattice matched wetting layer made of InAs in between QDs and substrate layer as shown in fig (1), which is experimentally viable, which can be also extended as a graded wetting layer. In order to minimize this problem, we explored QDs shape factor effects on induced stresses, which is partially relieved by the InAs pyramidal QDs. Already there are many attempts of insertion of QDs in single junction PVc to make the QDs based intermediate band solar cell (IBSC) for higher energy conversion efficiency and maximum absorption of the solar spectrum. Theoretical predictions for conversion efficiency of the IBSCs (also known as ladder approach or multi-step) are near 63% by making the use of InAs and GaAs QDs [40,41]. Our approach is to insert the QDs layer as IB into the current limiting junction of a multijunction solar cell to extend global efficiency. From previous work, we observed that the substrate layer made of AlAs gives fewer strain effects in structure compared to the GaAs substrate layer. To limit the final cost of solar cell it is desirable to produce PVc as thinner as possible. In multi-layered PVc, the best way to reduce the thickness of cell is to use the direct band gap materials which are also responsible for better absorption coefficient. Thus, the suggested PVc materials apparently have the desired properties for QDs based intermediate band multijunction PVc with maximum absorption of the solar spectrum. Different parameters for used materials are summarized in table (2).

Tab. 2: Material Parameters

| Materials | Band Gap (eV) | Electron Mobility (cm$^2$/V*s) | Lattice constant (nm) | Electron Effective Mass/$m_0$ |
|---|---|---|---|---|
| GaAs | 1.441 | 9000 | 0.5653 | 0.067 |
| InAs | 0.354 | 40000 | 0.605 | 0.027 |
| AlAs | 2.12 | 200 | 0.5660 | 0.1 |

### IV. Results and Discussion

In this section, all required parameters, simulation methodology for the QDs based multi-layered PVc to possess a heterogeneous type-II band structure with the effect of strain are discussed. Then, the properties of the IB and epitaxial growth of QDs are analyzed. 3D wavefunctions and absorption graphs are plotted for single QD. Finally, the potential of the proposed structure to form an IBSC is investigated. We explored a variety of materials and shapes of QDs, studied the system with two simulation setup models, first, two bands effective mass model, second, the 10-band tight-binding (TB) model. Obtained results through both models used for investigation of the confined electronic structure of InAs QD and optical properties.

In simulation, number of conduction and valence band is 4, which is the number of eigen energies to be solved in the CB and VB. The calculation methodology for both the models is such as, the two bands effective mass model uses the bottom of the CB and top of VB whereas the ten band TB model uses ten bands to characterize the materials and hence it is predicted that TB model should be more accurate. In the two-band model, it is assumed that the crystal structure of a material is simple cubic while TB model assumes the zincblende crystal structure. Strain simulation for the proposed device is performed with the TB model before calculating the eigenstates hence the induced strain in the intermediate band causes a significant change in eigenstates of QDs. The seed value is set to be 1000, it uses to calculate alloy distribution in the capping layer. Fermi level value is taken with respect to lowest eigen energy. In this study, Fermi level is defined as the position of the Fermi energy relative to the lowest energy state. Through trial and error, we concluded that Fermi level value 0.8ev gives suitable results. For absorption, Lorentzian function is used which can be expressed as follows,

$$L(x) = \frac{1}{\pi} \frac{0.5\Gamma}{(x-x_0)^2 + (0.5\Gamma)^2} \quad (8)$$

where $x_0$ is the center value and Gamma ($\Gamma$) is a parameter specifying the width. In our study, the value of Gamma is taken as 0.01, which decide the state broadening. TB model is implemented with spin orbit coupling effects, which is a relativistic particle's spin with its motion inside a confined potential space. Spin-orbit coupling plays a significant role in modifying energy levels. Configuration of falling solar light rays on QD and polarization plane with respect to horizontal and vertical planes is shown in fig. (2). $\theta$ is the angle between axis perpendicular to polarization plane and z-axis, and angle $\emptyset$ is in X-Y plane.

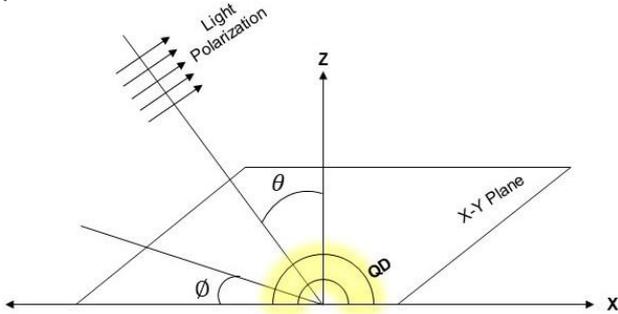

Fig. 2: Light Polarization Plane

A finite element approach has been adopted to solve the Schrödinger equations, considering that accurate analytical solutions for the Schrödinger equations are not possible for our system. In FEM analysis of the system maximum iteration is 15000. Convergence limit is taken as 1e-14, with the resolution of 1e-6. The number of eigen values for CB and VB is 4. For calculation of optical properties, the optical solver is used with matrix element type and atomic structure domain. All study has been done in normal temperature 300k [42]. Using the above-mentioned simulation methodology and numerical techniques we have plotted 3D wavefunction at different eight energy states which shows the successive quantum dot growth and electronic confinement. Resulting in the number of excitons (electrons-holes) produced per photon and photon absorption graphs have been plotted as a function of energy (eV) for single QD where InAs quantum dot serves mainly as the light-absorbing component. The fig. (3) and (4) shows the electronic confinement simulation of the 8 states energy level inside the pyramid InAs QD.

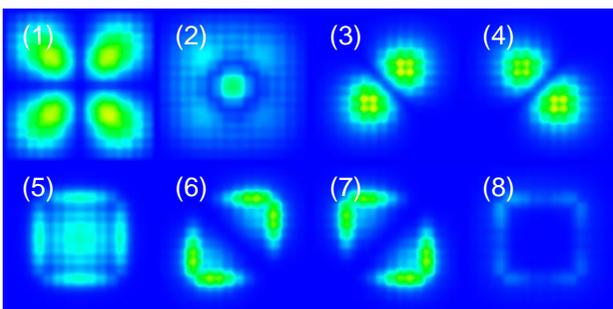

Fig. 3: 3D wavefunctions of the proposed InAs quantum dot inside an IB solar cell using two bands effective mass model at different energy levels from one to eight.

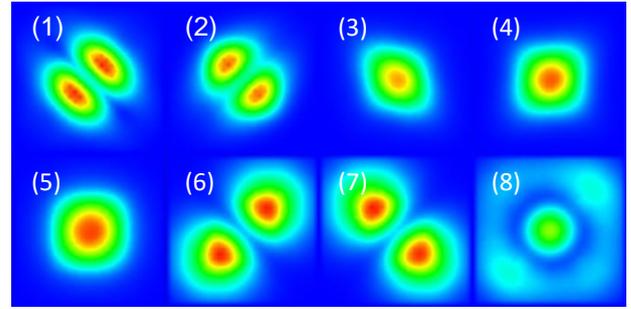

Fig. 4: 3D Wavefunctions of the proposed InAs quantum dot inside an IB solar cell using ten bands tight binding model at different energy levels from one to eight.

The 3D wavefunction of one to eight energy states of InAs QD are plotted in Fig. (3&4) as a function of QD energy levels. Fig. (3) is obtained by using the two bands effective mass model and fig. (4) is obtained by using the ten bands tight binding model. For both models, photocarrier electronic confinement inside pyramid QD is the lowest in 8th energy state. It can be observed that transition energy and electronic confinement calculation at each energy state using the tight binding model gives more accurate result compared to the two bands effective mass model. These results can be also used in deciding the spacing parameters between two quantum dots in a layer. Overlapping of the wavefunction of two successive elements in bulk nanocrystals always causes the resistance in producing the charge carriers efficiently. Confinement effects can be also observed in the VB and CB alignment at hetero-interfaces. Quantum confinement becomes more important for the characterization of carrier dynamics in the case of intra as well as inter-band tunneling phenomena.

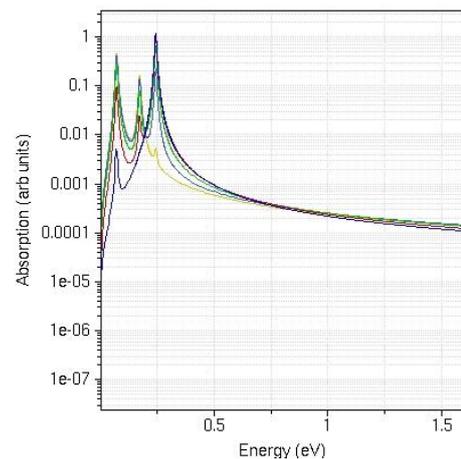

Fig. 5: Absorption graph for InAs QD as function of angle $\theta$ using effective mass model

Fig. (5), shows absorption results for InAs QD as a function of energy and angle $\theta$, using the effective mass model. We explored absorption factor for angle 0, 22.5, 45, 67.5 and 90 degree. As figure (5) suggest, after 0.5eV energy absorption coefficient is about the same and the peak value is around at 0.37eV, highest absorption is obtained at $\theta$ = 90 angles.

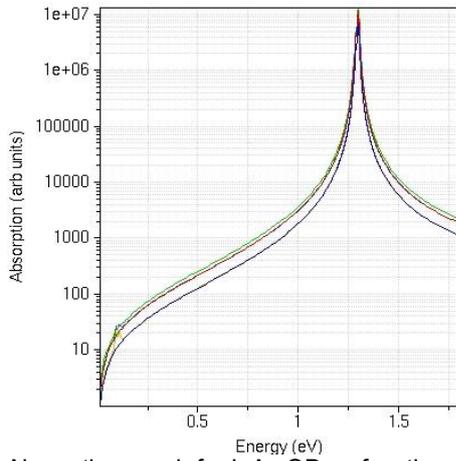

Fig. 6: Absorption graph for InAs QD as function of angle θ using tight binding model

Fig. (6) has been plotted using the tight binding model which gives the highest absorption coefficient at θ = 90 and 1.25eV energy value. So, this time also tight binding model gives better results.

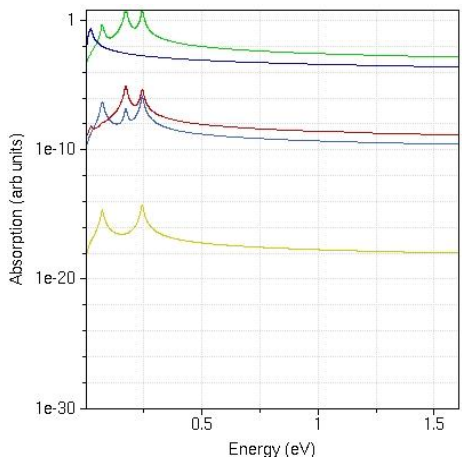

Fig. 7: Absorption graph for InAs QD as function of energy (eV) using tight binding model

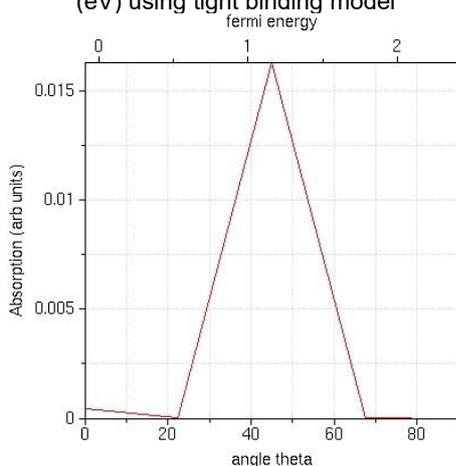

Fig. 8: Local absorption graph for InAs QD as function of energy (eV) and angle θ using tight binding model

Figure (7) summarize the absorption coefficient for different Fermi energy level values while keeping the constant value of angle θ. Min absorption shows the yellow line and the maximum is shown by the green line. Fermi energy level values for both is 2.5eV and 1.25eV respectively. Other intermediate lines are plotted for 1.87, 0.625 and 0eV. Figure (8) has been plotted as a function of Fermi energy and angle θ, from this graph we can conclude that maximum absorption achieved with Fermi energy value of 1.25eV with angle θ value of 45 degree, which is a real scenario.

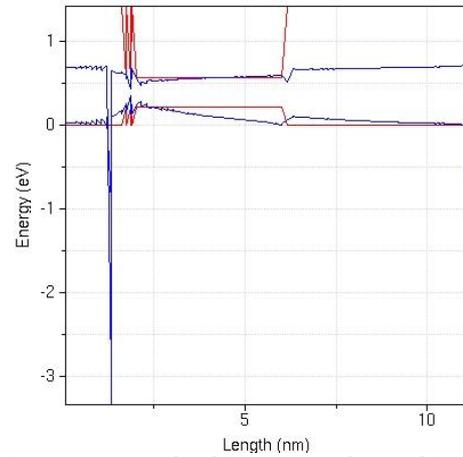

Fig. 9: Band structure for CB and VB of InAs QD including strain effect in system using k.p. model

In order to search material combinations for QD based multijunction PVc, we have calculated the strain profile in and around the InAs QD due to lattice mismatch and heated charge carriers between the barrier and QDs material. Plotted the graph of strain effect on the CB and VB band diagram. Fig. (9) shows the band bending before and after strain effects. Red lines show CB and VB without considering the effects of strain while blue lines summarize the effect of strain on CB and VB. Long well in the blue line in CB is due to calculation errors. From this, we can conclude that still there is a large effect of strain in band bending and structure which can cause significant changes in bulk nanocrystals of PVc.

## V. Conclusion and Recommendations

The study on quantum dots based heterostructure multijunction photovoltaics cell have been presented using quantum dot lab (an initiative by National Science Foundation). Hurdles faced by PVc researchers and the basic principle of the solar cell have been reviewed. Parameters which decide the size of QD and electronic confinement investigated extensively. The complete architecture of PVc with single QD was presented and 3D wavefunction for InAs QD has been plotted using two models, one, two bands effective mass model, other, ten bands tight binding model. The effects of induced strain on band structure has shown. Dimensions of QD has been proposed and found that height of 3nm is appropriate. From this study we can conclude three major points, first, the tight binding model gives better result in the approximate numerical calculation of PVc compared to an effective mass model. Second, ideally, the angle θ value 90 degree gives maximum absorption which is not in a real situation because most of the day time solar panels are inclined and not facing sun ray's in perpendicular direction. Third, the introduction of the wetting layer between the barrier layer and QDs array is the best solution for lattice mismatch and induced strain. And the substrate layer made of AlAs is suggested.

The challenge is now to further improvement in QDs array fabrication and making the use of it in the multijunction PVc efficiently as well as improvement in QDs materials for better optical and electrical performance. Till now somehow, we can say that QDs based solar cells are well explored and have enough data for device fabrication but still structure and composition of one-dimensional confined quantum wells and two-dimensional confined quantum wires is less explored. As previous work suggests that already quantum weirs-based laser and other optoelectronics devices have been constructed successfully and available commercially. In this paper, we haven't discussed about recombination effect which is the major problems in present PVc. Recombination factor plays a significant role to decide how thick efficient PVc can be made and gives information about the diffusive length and charge mobility. Still, we need to investigate and study about QDs and their use in photo technology because quantum dos are more awesome as we hoped. In future new advanced measurement techniques may give a clear picture of this promising technology and may finally dissolve doubts about the potential of QDs. Improvement in Nanomachining and manufacturing process is also prime need to make this success and for producing coherent tiny quantum dots with an easy and economical way. In next upcoming decade, we just need a more compact and efficient instrument for making the next generation medical devices, artificial intelligence, quantum computers, cryptography and semiconductors etc., and quantum dots can play key role for deploying these things.

## Acknowledgements

This work has been done independently. The authors thank to Bose.X research group members for fruitful discussions regarding quantum confinement and materials.

## Nomenclature

$\lambda_{e/h}$: de Broglie wavelength (nm)
$h$ : Plank Constant (Joule-seconds)
ε: dielectric constant
$R_b$: Bohr Radius (nm)
E: Energy (eV)
V: Potential (Volts)
X: Position Vector
p: Momentum Operator
Greek letters
μ: Reduced eff mass of the electron and hole
$\Psi_{n,k}$: Wavefunction
$\Gamma$: Gamma
$\emptyset, \theta$: Angle (degree)
Subscripts,
eff: Effective
e: Electron
h: Hole

## References

[1]. Basov, D. N., Averitt, R. D., van der Marel, D., Dressel, M. & Haule, K. Electrodynamics of correlated electron materials. Rev. Mod. Phys. 83, 471–541 (2011).

[2]. Inoue, J. & Tanaka, A. Photoinduced transition between conventional and topological insulators in two-dimensional electronic systems. Phys. Rev. Lett. 105, 017401 (2010).

[3]. Basov, D. N., Averitt, R. D. & Hsieh, D. Controlling the properties of quantum materials. Nat. Mater. 16, 1077–1088 (2017).

[4]. Tokura, Y., Kawasaki, M. & Nagaosa, N. Emergent functions of quantum materials. Nat. Phys. 13, 1056–1068 (2017).

[5]. Brus L (1986) Electronic wave functions in semiconductor clusters: experiment and theory. J Phys Chem 90(12):2555–2560.

[6]. Canham LT (1990) Silicon quantum wire array fabrication by electrochemical and chemical dissolution of wafers. Appl Phys Lett 57:1046–1048.

[7]. Shen JH, Zhu YH, Yang XL, Li CZ (2012) Graphene quantum dots: emergent nanolights for bioimaging, sensors, catalysis and photovoltaic devices. Chem Commun 48:3686–3699.

[8]. Ding ZF, Quinn BM, Haram SK, Pell LE, Korgel BA, Bard AJ (2002) Electrochemistry and electrogenerated chemiluminescence from silicon nanocrystal quantum dots. Science 296:1293–1297.

[9]. Burda, C., Chen, X., Narayanan, R., El-Sayed, M.A.: Chemistry and properties of nanocrystals of different shapes. Chem. Rev. 105, 1025–1102 (2005).

[10]. Mooradian A (1969) Photoluminescence of metals. Phys Rev Lett 22(5):185–187

[11]. Maurer, P. C. et al. Room-temperature quantum bit memory exceeding one second. Science 336, 1283–1286 (2012).

[12]. Emin, S., Singh, S.P., Han, L., Satoh, N., Islam, A.: Colloidal quantum dot solar cells. Solar Energy 85, 1264–1282 (2011).

[13]. J. Kusuma, R. Geetha Balakrishna, A review on electrical characterization techniques performed to study the device performance of quantum dot sensitized solar cells, Solar Energy 159 (2018) 682–696.

[14]. Adla, A., 2010. Instrumentation for Quantum Efficiency Measurement of Solar Cells. Photovoltaics World, pp. 28–31 July/August.

[15]. Saga T. Advances in crystalline silicon solar cell technology for industrial mass production. NPG Asia Mater 2010; 2, 96–102.

[16]. Nozik, A.J.: Quantum dot solar cell. Physica E 14, 115–120 (2002).

[17]. O'Regan, B., Gra¨tzel, M.: A low cost, high-efficiency solar based on dye-sensitized colloidal TiO2 films. Nature 353, 737–740 (1991).

[18]. Gra¨tzel, M.: Solar energy conversion by dye-sensitized photovoltaic cells. Inorg. Chem. 44, 6841–6851 (2005).

[19]. Parkinson, B.A.: Multiple exciton collection in a sensitized photovoltaic system. Science 330, 63 (2010).

[20]. Shah, A., Torres, P., Tscharner, R., Wyrsch, N., Keppner, H.: Photovoltaic technology: the case for thin-film solar cells. Science 285, 692–698 (1999).

[21]. Gime´nez, S., Mora-Sero´, I., Macor, L., Guijarro, N., Lana-Villarreal, T., Go´mez, R., Diguna, L.J., Shen, Q., Toyoda, T., Bisquert, J.: Improving the performance of

colloidal quantum-dot sensitized solar cells. Nanotechnology 20, 295204 (2009).

[22]. Kongkanand, A., Tvrdy, K., Takechi, K., Kuno, M., Kamat, P.V.: Quantum dot solar cells. Tuning photo response through size and shape control of CdSe–TiO2 architecture. J. Am. Chem. Soc. 130, 4007–4015 (2008).

[23]. Murray CB, Norris DJ, Bawendi MG. Synthesis and characterization of nearly monodisperse CdE (E = sulfur, selenium, tellurium) semiconductor Nano crystallites. J Am Chem Soc. 1993,115, 8706–8715.

[24]. Mogi, M. et al. A magnetic heterostructure of topological insulators, a candidate for axion insulator. Nat. Mater. 16, 516–521 (2017).

[25]. Ohtomo, A. & Hwang, H. Y. A high-mobility electron gas at the LaAlO3/SrTiO3 heterointerface. Nature 427, 423–426 (2004).

[26]. Boschker, H. & Mannhart, J. Quantum-matter heterostructures. Annu. Rev. Condens. Matter Phys. 8, 145–164 (2017).

[27]. Novoselov, K. S., Mishchenko, A., Carvalho, A. & Castro Neto, A. H. 2D materials and van der Waals heterostructures. Science 353, aac9439 (2016).

[28]. Ankul Prajapati et al. Numerical Investigations on the Type-II Band Alignment and Quantum Efficiency of Multijunction Solar Cell using Anderson's Rule, Third ISEES International Conference on Sustainable Energy and Environmental Challenges (III-SEEC) proceeding, 3 (2018) 424-426.

[29]. Zhong, D. et al. Van der Waals engineering of ferromagnetic semiconductor heterostructures for spin and valleytronics. Sci. Adv. 3, e1603113 (2017).

[30]. Byoung Don Kong, Zhenghe Jin, and Ki Wook Kim, Hot-Electron Transistors for Terahertz Operation Based on Two-Dimensional Crystal Heterostructures, Physical Review Applied 2, 054006 (2014).

[31]. Octavi E. Semonin, Joseph M. Luther, and Matthew C. Beard, Quantum dots for next-generation photovoltaics, (2012) VOLUME 15 materials-today.

[32]. Nozik, A. J., Physica E (2002) 14, 115–120.

[33]. Murphy, J. E., Beard, M. C., and Nozik, A. J., J Phys Chem B (2006) 110, 25455–25461.

[34]. E. O. Kane, J. Phys. Chem. Solids 1, 249 (1957).

[35]. E. O. Kane, Semicond. Semimetals 1, 75 (1966).

[36]. J. M. Luttinger and W. Kohn, Phys. Rev. 97, 869 (1955).

[37]. Stanko Tomić Physical Review B 82, 195321 (2010).

[38]. M. Cardona and F. H. Pollak, Phys. Rev. 142, 530 (1966).

[39]. S. Richard, F. Aniel, and G. Fishman, Phys. Rev. B 70, 235204 (2004).

[40]. 3A. Marti, L. Cuadra, and A. Luque, IEEE Trans. Electron Devices 49, 1632 (2002).

[41]. S. M. Hubbard, C. D. Cress, C. G. Bailey, R. P. Raffaelle, S. G. Bailey et al., Appl. Phys. Lett. 92, 123512 (2008).

[42]. Building and Deploying Community Nanotechnology Software Tools on nanoHUB.org and Atomistic simulations of multimillion-atom quantum dot nanostructures", Gerhard Klimeck, Marek Korkusinski, Haiying Xu, Seungwon Lee, Sebastien Goasguen and Faisal Saied, Proceedings of the 5th IEEE Conference on Nanotechnology, 2: pg. 807, 07 (2005).